\documentclass[showpacs,amssymb,aps,twocolumn]{revtex4}
\usepackage{amsmath}
\usepackage{amstext}
\usepackage{amsopn}
\usepackage{amsfonts}
\usepackage{amssymb}
\usepackage{bbm}
\usepackage{accents}
\usepackage{empheq}
\usepackage{graphicx}
\usepackage{epsf}
\usepackage{graphics}
\usepackage[latin1]{inputenc}

\begin{document}

\title{Thermal effective action for $\mathbf{1+1}$ dimensional massive QED}

\author{Ashok Das$^{a,b}$ and J. Frenkel$^{c}$\footnote{$\ $ e-mail: das@pas.rochester.edu,  
jfrenkel@fma.if.usp.br}}
\affiliation{$^a$ Department of Physics and Astronomy, University of Rochester, Rochester, 
NY 14627-0171, USA}
\affiliation{$^b$ Saha Institute of Nuclear Physics, 1/AF Bidhannagar, Calcutta 700064, India}
\affiliation{$^{c}$ Instituto de Física, Universidade de São Paulo, 05508-090, São Paulo, SP, BRAZIL}

\begin{abstract}
In continuation of our earlier proposal \cite{dasfrenkel, dasfrenkel1} for evaluating thermal effective actions, we determine the exact fermion propagator in $1+1$ dimensional massive QED. This  propagator is used to derive the finite temperature effective action of the theory which generates systematically all the one loop Feynman amplitudes calculated in thermal perturbation theory. Various aspects of the effective action including its imaginary part are discussed. 
\end{abstract}


\maketitle
\newpage

\section{Introduction}

At zero temperature the effective action for a system of fermions interacting with a background field, which incorporates all the one loop corrections in the  theory, can be  beautifully derived by a method due to Schwinger \cite{schwinger} known as the proper time formalism. It naturally introduces a gauge invariant regularization (in the case of gauge backgrounds) to regularize the ultraviolet divergences that arise at zero temperature. For example, we note that the effective action for  a fermion with mass $m$  interacting with a background gauge field is given by (for space-time dimensions $d\geq 2$)
\begin{equation}
\Gamma_{\rm eff} [A] = - i {\rm Tr}\, \ln (\gamma^{\mu} (i\partial_{\mu} -  g A_{\mu}) - m) = - i {\rm Tr}\, \ln H,\label{effaction}
\end{equation}
where $A_{\mu}$ denotes the background field and $g$ represents the coupling to the background and 
we have identified
\begin{equation}
H =  \gamma^{\mu} (i\partial_{\mu} -  g A_{\mu}) - m.
\end{equation}
Here ``Tr" denotes trace over a complete basis as well as the trace over Dirac indices. (The method applies equally well to a scalar background, but for the purpose of our discussions we have specialized to a gauge field background.)  Schwinger expressed the effective action \eqref{effaction} in a regularized integral form as 
\begin{equation}
\Gamma_{\rm eff} [A] = \lim_{\nu\rightarrow 0}\ i \int\limits_{0}^{\infty} \frac{d\tau}{\tau^{1-\nu}}\,
{\rm Tr}\,e^{-\tau H},\label{effaction1}
\end{equation}
where $\tau$ is known as the ``proper time" parameter. The idea here is that the operator $e^{-\tau 
H}$ in the integrand can be thought of as the  evolution operator for the Euclidean time $\tau$ with $H$ 
denoting the (Hamiltonian) generator for the evolution. (The ``proper time" can also be made 
Minkowskian with appropriate $i\epsilon$ prescription.) As a result, we can write the proper time 
evolution equations as 
\begin{align}
\frac{dx^{\mu}}{d\tau}  & = -i [x^{\mu}, H],\notag\\
\frac{dp_{\mu}}{d\tau}  & = -i [p_{\mu}, H].\label{dynamicaleqn}
\end{align}
If these equations can be solved and $x^{\mu}(\tau)$ (or $p_{\mu} (\tau)$) can be determined in a 
closed form, then one can evaluate the trace in \eqref{effaction1} in the eigenbasis 
$|x^{\mu} (\tau)\rangle$ (or $|p_{\mu}(\tau)\rangle$) and evaluate the (gauge invariant) regularized 
effective action in a closed 
form as well (or at least give an integral representation). In the case of fermions interacting with a constant background electromagnetic field, this has been profitably used to 
calculate the imaginary part of the effective action which describes the decay rate of the vacuum \cite{schwinger}. However, solving the dynamical equations in \eqref{dynamicaleqn} is, in general, not easy when nontrivial interactions are present. When the dynamical equations cannot be solved in a closed form, the method due to Schwinger leads to a perturbative determination of the effective action.

In the past couple of decades, there have been several attempts \cite{generalization,frenkel} to 
generalize the method due to Schwinger to finite temperature \cite{temp,das} and to determine the 
imaginary part of the effective action, leading to conflicting results \cite{generalization}. In 
\cite{dasfrenkel, dasfrenkel1} we have presented an alternative method for determining finite temperature effective actions for fermions interacting with an arbitrary background field. We believe that since the amplitudes at finite temperature are ultraviolet finite, unlike those at zero temperature, it is not necessary to generalize the method due to Schwinger to finite temperature. After all, the proper time method was designed to provide a (gauge invariant) ultraviolet regularization which is not necessary at finite temperature. Therefore, we have proposed \cite{dasfrenkel, dasfrenkel1} a direct method for evaluating finite temperature effective actions based mainly on the general properties of systems at finite temperature. As applications of our method, we have determined the complete thermal effective actions for the $0+1$ dimensional QED as well as the $1+1$ dimensional Schwinger model (massless QED) in \cite{dasfrenkel, dasfrenkel1}. 

In this paper we extend our calculation of thermal effective actions to the case of $1+1$ dimensional massive QED. This model is significant from various points of view. For example, any realistic physical model of interest would involve massive fermions. Secondly, in the $1+1$ dimensional Schwinger model (massless QED), the left and the right handed fermions decouple and propagate on the light-cone  giving the model very broadly (not exactly though) a $0+1$ dimensional character. A mass for the fermion, on the other hand, would couple the two modes and the behavior will not be as simple. Furthermore, in the models in $3+1$ dimensions with massive fermions in a background electric field which are known to have closed form effective actions at zero temperature \cite{soluble, soluble1}, the background electric field generally points in a fixed direction with either a (special) time dependence or a (special) dependence on only one of the space coordinates. Therefore, these models are effectively $1+1$ dimensional models of massive QED. (The case of magnetic field backgrounds would require effectively the study of $2+1$ dimensional theories.) 

One of the interests in studying finite temperature effective actions is to determine the effect of temperature on the imaginary part of the action which is related to the vacuum decay rate. At zero temperature, for example, this is thought of as arising due to pair creation. On the other hand, at finite temperature we know that there are additional channels of reaction possible. The $1+1$ dimensional (massive) QED, in fact, provides an excellent model to study not only the temperature dependence of the imaginary part of the effective action, but possibly also in identifying the processes responsible for destabilizing the vacuum at finite temperature. The additional thermal processes are known to lead to nontrivial non-analytic behavior in amplitudes at finite temperature and the effect of this in the effective action can also be studied explicitly in this model.

The present paper is organized as follows. In section {\bf II} we recapitulate our proposal 
\cite{dasfrenkel, dasfrenkel1} for evaluating effective actions at finite temperature and also summarize the essential results obtained for the Schwinger model (massless QED in $1+1$ dimension). In section {\bf III} we determine the massive fermion propagator in the massive QED in terms of the complete massless propagator and describe various of its properties. The temperature dependent effective action for  the $1+1$ dimensional massive QED is also derived in this section where we discuss its imaginary part as well. In section {\bf IV} we calculate explicitly the quadratic effective action at finite temperature to support the general discussion of the behavior of the effective action. We present a brief summary of our results  in section {\bf V}.

\section{Proposal} 

From the definition of the effective action \eqref{effaction} for a system of massive fermions 
interacting with an arbitrary gauge field background, it is straightforward to obtain   
\begin{equation}
\frac{\partial\Gamma_{\rm eff}}{\partial m} = \int dt d\mathbf{x}\, {\rm tr}\,S (t,\mathbf{x}; t, \mathbf{x}) = {\rm Tr}\,S,
\label{propagator0}
\end{equation}
where $S (t,\mathbf{x};t', \mathbf{x}')$ denotes the complete Feynman propagator for the fermion 
(including the factor $i$) in the presence of the background field, ``tr" stands for trace over the Dirac indices and ``Tr" denotes the trace over a complete basis as well as the Dirac trace. If the fermion does not have a mass (as in the Schwinger model \cite{schwinger2}), we note that the variation of the effective action with respect to the background gauge field leads to 
\begin{equation}
\frac{\delta \Gamma_{\rm eff}}{\delta A_{\mu} (t,\mathbf{x})} =  g\, {\rm tr}\, \left(\gamma^{\mu} S (t, \mathbf{x}; t, \mathbf{x})\right).
\label{propagator1}
\end{equation}
In either case, we note that it is the fermion propagator that is relevant in \eqref{propagator0} or \eqref{propagator1} for the evaluation of the effective action. 

As we have emphasized in \cite{dasfrenkel, dasfrenkel1},  the real time formalism \cite{das} (we use the closed time path formalism due to Schwinger \cite{das, schwinger1}) is more suited for the evaluation of the effective action at finite temperature. Furthermore, as we have pointed out earlier \cite{das,tor}, the real time calculations can be carried out quite easily in the mixed space where the 
spatial coordinates $\mathbf{x}$ have been Fourier transformed. We note that in the mixed space we can write \eqref{propagator0} and \eqref{propagator1} as
\begin{align}
\frac{\partial\Gamma_{\rm eff}}{\partial m} & = {\rm Tr}\,S = \int dt d\mathbf{p}\, {\rm tr}\,S (t,\mathbf{p}; t, \mathbf{-p}),\nonumber\\
\frac{\delta\Gamma_{\rm eff}}{\delta A_{\mu} (t, -\mathbf{p})} & = g \int d\mathbf{k}\, {\rm tr}\left(\gamma^{\mu} S (t, \mathbf{k+p}; t, \mathbf{k})\right).\label{propagator}
\end{align} 

Since the effective action is so intimately connected with the fermion propagator, our proposal is to 
determine the complete fermion propagator at finite temperature directly such that 
\begin{enumerate}
\item[(i)] it satisfies the appropriate equations for the complete propagator of the theory, 
\item[(ii)] it satisfies the necessary symmetry properties of the theory such as the Ward identity, 
\item[(iii)] and most importantly, it satisfies the anti-periodicity property associated with a finite 
temperature fermion propagator \cite{das}. 
\end{enumerate}
In fact, it is the third requirement that is quite important in a direct determination of the propagator. 
We note that this last condition is missing at zero temperature which makes it difficult to determine 
the complete propagator. When the theory has no ultraviolet divergence (so that it does not need a regularization at zero temperature), this exact fermion propagator of the theory would lead to the complete effective action including the correct zero temperature part. On the other hand, if the theory needs to be regularized at zero temperature, this propagator will not yield the correct zero temperature effective action. However, we note that our interest is in the finite temperature part of the effective action which does not need to be regularized (it is not ultraviolet divergent) and will be determined correctly in this approach. 

In \cite{dasfrenkel, dasfrenkel1} we determined the complete fermion propagator at finite temperature for $0+1$ dimensional QED as well as for the $1+1$ dimensional Schwinger model \cite{schwinger2} and this led to the thermal effective actions for these theories. Here we recapitulate briefly the results for the Schwinger model for later use. The Schwinger model describes massless QED in $1+1$ dimensions. The masslessness of the fermion leads to a separation of the theory into two decoupled sectors of right handed and left handed fermions moving on the light-cone and coupling only to the light-cone components $A_{\pm}$ of the gauge field. Therefore, the effective action can be naively expected to be a sum of two decoupled terms, one depending only on $A_{+}$ while the other depending on $A_{-}$. This expectation is indeed almost correct except for the axial anomaly which arises at zero temperature due to the ultraviolet divergence in the theory and leads to a quadratic term in the effective action involving $A_{+}A_{-}$. However, at finite temperature there is no ultraviolet divergence and we expect the finite temperature effective action to be decoupled in $A_{+}$ and $A_{-}$. Therefore, for the purposes of finite temperature effective action, we can study the decoupled right handed and the left handed fermion sectors separately.

In the closed time path formalism, the complete contour ordered (thermal) propagators for the right handed and left handed fermions can be determined exactly. For the right handed fermion, for example, we can Fourier transform the light-cone coordinate $x^{-}$ (so that we are in the mixed space) and introduce the operators $\hat{A}_{+ c} (x^{+})$ and $\hat{S}_{R c}$ such that (see \cite{dasfrenkel, dasfrenkel1} for details, $c$ stands for quantities defined on the contour in the closed time path formalism)
\begin{align}
S_{R c} (x^{+},x'^{+};k+p,k) & =  \langle k+p|\hat{S}_{R c} (x^{+},x'^{+})|k\rangle,\nonumber\\ 
A_{+ c} (x^{+},p-k) & =  \langle p|\hat{A}_{+ c} (x^{+})|k\rangle,\label{operators}
\end{align}
where $k=k_{-}, p=p_{-}$, namely, the momenta conjugate to $x^{-}$. With this notation, the complete contour ordered propagator for the right handed fermion has the form
\begin{eqnarray}
\hat{S}_{R c} (x^{+},x'^{+})  & = & \frac{1}{4}\, e^{-ie \int d \bar{x}^{+}\,\theta_{c} (x^{+}-
\bar{x}^{+}) \hat{A}_{+c} (\bar{x}^{+})}\nonumber\\
& & \times \big({\rm sgn}_{c} (x^{+}-x'^{+}) + 1 - 2(\hat{\cal O}_{+} + 1)^{-1}\big)\nonumber\\
& & \times e^{ie \int d \bar{x}^{+}\,\theta_{c} (x'^{+}-\bar{x}^{+}) \hat{A}_{+c} (\bar{x}^{+})},
\label{1plus1propagator}
 \end{eqnarray}
where $\theta_{c}$ and ${\rm sgn}_{c}$ denote the step function and the alternating step function on the contour with
\begin{equation}
{\rm sgn}_{c} (x^{+}-x'^{+}) = \theta_{c} (x^{+}-x'^{+}) - \theta_{c}(x'^{+}-x^{+}),
\end{equation}
and $\hat{\cal O}_{+}$ which contains all the nontrivial information about interactions and temperature is 
independent of the coordinates $x^{+},x'^{+}$ and is given by
\begin{equation}
 \hat{\cal O}_{+}  = e^{\frac{ie (\hat{a}_{+ (+)}-\hat{a}_{+ (-)})}{2}}\,e^{\frac{\beta \hat{K}}{2}}
\,e^{\frac{ie (\hat{a}_{+(+)}-\hat{a}_{+(-)})}{2}}.\label{operatorO}
\end{equation}
Here $\hat{K}$ denotes the momentum operator satisfying
\begin{equation}
\hat{K} |p\rangle = p |p\rangle,
\end{equation}
and ($(\pm)$ with the parenthesis denote the thermal 
indices while $+$ without the parenthesis represents the light-cone component of the background 
field)
\begin{equation}
\hat{a}_{+ (\pm)} = \int\limits_{-\infty}^{\infty} d x^{+}\,\hat{A}_{+(\pm)} (x^{+}).
\end{equation}
It can be checked that the propagator \eqref{1plus1propagator} satisfies the equation for the Green's function (with a factor of ``$i$"), the correct Ward identities as well as the anti-periodicity necessary for a thermal fermion propagator. Therefore, it satisfies all the requirements of our proposal. Furthermore, this also satisfies the Lippmann-Schwinger equation \cite{lippman} to all orders so that it agrees with the perturbative expansion order by order. The effective action for this sector can now be determined using \eqref{propagator} and leads to
\begin{equation}
\Gamma_{R\, {\rm  eff}}  = - \frac{i}{2} \int \frac{dk}{2\pi}\,\langle k| \ln \cosh (\frac{1}{2} \ln \hat{\cal O}_{+}) - 
\ln \cosh \frac{\beta\hat{K}}{4}|k\rangle.\label{raction}
\end{equation}
The complete contour ordered propagator as well as the effective action for the left handed fermion has forms similar to \eqref{1plus1propagator} and \eqref{raction} with $A_{+}\rightarrow A_{-}, x^{+}\rightarrow x^{-}, (k_{-},p_{-})\rightarrow (k_{+},p_{+})$. These effective actions coincide with the perturbative determination of the thermal effective action \cite{adilson} order by order. We note here that the contour ordered propagators take a $2\times 2$ matrix structure (in the thermal space) when we restrict the time arguments to the two (``$\pm$") branches of the contour in the closed time path formalism.

 \section{Massive QED in $\mathbf{1+1}$ dimensions}

With the background of results for the Schwinger model in the last section, we are ready to study the question of the thermal effective action for massive QED in $1+1$ dimensions. The theory is described by the Lagrangian density
\begin{equation}
{\cal L} = \overline{\psi} (t,x) \left(\gamma^{\mu} (i\partial_{\mu} - eA_{\mu} (t,x)) - m\right) \psi  (t,x),
\label{QEDL}
\end{equation}
where $m$ denotes the mass of the fermion and $e$ the electric charge (or the coupling to the background). We use the Bjorken-Drell metric with signatures $(+,-)$, $\mu=0,1$ and
\begin{equation}
\gamma^{0} = \sigma_{1},\quad \gamma^{1} = -i\sigma_{2},\quad \gamma_{5} = \gamma^{0}\gamma^{1} = \sigma_{3},\label{diracmatrices}
\end{equation}
where $\sigma_{i}, i=1,2,3$ denote the three Pauli matrices. As we have pointed out earlier, in the absence of the mass term for the fermion, the Lagrangian density \eqref{QEDL} decomposes into a sum of two (decoupled) terms involving right handed and left handed fermions interacting with light-cone components of the background field and propagating on the light-cone. The mass term, on the other hand, couples these two modes and, therefore, the propagation ceases to be on the light-cone in the massive theory.

\subsection{Massive propagator}

\begin{figure}[ht!]
\begin{center}
\includegraphics{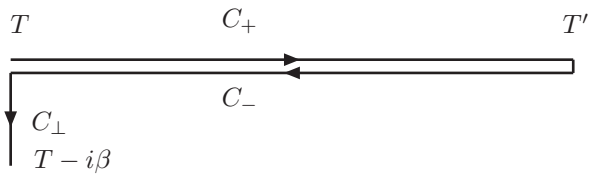}
\end{center}
\caption{The closed time path contour in the complex $t$-plane. Here
  $T\rightarrow -\infty$, while $T'\rightarrow \infty$ and $\beta$
  denotes the inverse temperature (in units of the Boltzmann constant
  $k$) \cite{das}.}
\label{1}
\end{figure}

In our method, the determination of the complete propagator of the theory is of crucial importance. So, in this section, let us study the propagator for the massive theory in detail. At zero temperature the propagator satisfies the equation
\begin{equation}
\left(\gamma^{\mu} (i\partial_{\mu} - e A_{\mu}) - m\right) S^{(m)} (x,x') = i \delta^{2} (x-x'),
\end{equation}
where $S^{(m)} (x,x')$ denotes the propagator for the massive theory (including the factor ``$i$"). At finite temperature the propagator can be described either as a contour ordered propagator defined on the contour in the complex $t$-plane as shown in Fig. \ref{1} or as a $2\times 2$ (thermal) matrix representing the four possible propagations involving the two real branches $C_{\pm}$ of the complex contour (see \cite{das} for details).
To make contact with perturbation theory (to be described in a later section) we  follow the matrix description (although everything can be discussed in terms of the complex contour equally well). Therefore, in this description, the fermion propagator is not only a $2\times 2$ matrix in the Dirac space, but also a $2\times 2$ matrix in the thermal space (the doubling of fields at finite temperature in the real time formalism arises in any dimensions, but the $2\times 2$ Dirac structure is special to $1+1$ dimensions). In the operator notation, the finite temperature (matrix) propagator satisfies the equation
\begin{equation}
\left(\gamma^{\mu} (i\partial_{\mu} - e \hat{A}_{\mu}) - m \tau\right) \hat{S}^{(m)}  = i \mathbbm{1},\label{matrixpropagator}
\end{equation}
where
\begin{equation}
\hat{A}_{\mu} = \begin{pmatrix}
\hat{A}_{\mu (+)} & 0\\
0 & - \hat{A}_{\mu (-)}
\end{pmatrix},\quad \tau = \begin{pmatrix}
1 & 0\\
0 & -1
\end{pmatrix},\label{thermalmatrices}
\end{equation}
denote thermal matrices representing the background as well as mass interactions on the $C_{\pm}$  branches of the contour. The negative sign on the $C_{-}$ branch reflects the fact that time is decreasing along this branch of the contour.

\begin{figure}[ht!]
\begin{center}
\includegraphics[scale=.7]{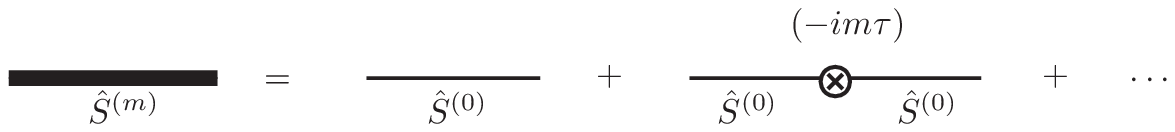}
\end{center}
\caption{The expansion of the massive propagator \eqref{lseqnexpansion} with $\otimes$ denoting a mass insertion $(-im\tau)$.}
\label{2}
\end{figure}

It follows from \eqref{matrixpropagator} that when $m=0$, the propagator satisfies the equation
\begin{equation}
\left(\gamma^{\mu} (i\partial_{\mu} - e \hat{A}_{\mu})\right) \hat{S}^{(0)}  = i \mathbbm{1}.\label{masslessmatrixpropagator}
\end{equation}
In terms of the complete massless propagators $\hat{S}_{R}$ and $\hat{S}_{L}$ determined in \cite{dasfrenkel, dasfrenkel1}, we note that we can write
\begin{equation}
\hat{S}^{(0)} = \begin{pmatrix}
0 & \hat{S}_{R}\\
\hat{S}_{L} & 0
\end{pmatrix},\label{explicitmasslesspropagator}
\end{equation}
where each element of the Dirac matrix is a $2\times 2$ thermal matrix. The off-diagonal nature of $\hat{S}^{(0)}$ (in the Dirac space) follows from the fact that while the massless propagator is defined as $\langle T (\psi\psi^{\dagger})\rangle$, the $m=0$ limit of \eqref{matrixpropagator} leads to $\langle T (\psi \overline{\psi})\rangle$ and the two differ by $\gamma^{0}$ which is off-diagonal (see \eqref{diracmatrices}). Using \eqref{masslessmatrixpropagator} in \eqref{matrixpropagator} leads to 
\begin{equation}
\left(i (\hat{S}^{(0)})^{-1} - m\tau\right) \hat{S}^{(m)} = i \mathbbm{1},
\end{equation}
which determines
\begin{align}
\hat{S}^{(m)} & = i\left(i (\hat{S}^{(0)})^{-1} - m\tau\right)^{-1} = \left(\mathbbm{1} + im \hat{S}^{(0)} \tau\right)^{-1} \!\!\hat{S}^{(0)}\nonumber\\
& = \left(\mathbbm{1} - \hat{S}^{(0)} (-im\tau)\right)^{-1} \!\!\hat{S}^{(0)}\nonumber\\
& = \hat{S}^{(0)}\left(\mathbbm{1} - (-im\tau)\hat{S}^{(0)}\right)^{-1}\!\!.\label{lseqn}
\end{align}
Equation \eqref{lseqn} which determines the massive propagator for the fermion can be thought of as a Lippmann-Schwinger equation \cite{lippman} where the interaction is a mass insertion (without any external field). However, the insertion vertex involves the thermal matrix $\tau$ and a matrix product of thermal matrices implies a sum over the thermal ``$\pm$" indices at the insertion vertex (just as there is summation over the Dirac indices at the vertex). Relation \eqref{lseqn} can be expanded to have the form
\begin{widetext}
\begin{equation}
\hat{S}^{(m)} = \sum_{n=0}^{\infty} \left(\hat{S}^{(0)} (-im\tau)\right)^{n} S^{(0)} = \hat{S}^{(0)} + \hat{S}^{(0)} (-im\tau) \hat{S}^{(0)} + \hat{S}^{(0)} (-im\tau) \hat{S}^{(0)} (-im\tau) \hat{S}^{(0)} + \cdots,\label{lseqnexpansion}
\end{equation}
\end{widetext}
which can be represented diagrammatically as shown in Fig. \ref{2}.
Since the massless propagator $\hat{S}^{(0)}$ satisfies the appropriate anti-periodicity conditions as well as the gauge Ward identities, it follows from \eqref{lseqn} or \eqref{lseqnexpansion} that $\hat{S}^{(m)}$ also satisfies these properties (this is crucial in our proposal and note that while \eqref{lseqn} or \eqref{lseqnexpansion} involve integration over intermediate points, only the initial and the final times are relevant for anti-periodicity)

To understand the meaning of \eqref{lseqn} or \eqref{lseqnexpansion}, we note that at zero temperature, if we denote the free propagators for a massive and a massless fermion as
\begin{equation}
S^{(m)} (p) = \frac{i}{p\!\!\!\slash - m},\quad S^{(0)} (p) = \frac{i}{p\!\!\!\slash},\label{freepropagator}
\end{equation}
then we can write
\begin{align}
S^{(m)} (p) & = i (p\!\!\!\slash - m)^{-1} = \left(\mathbbm{1} - S^{(0)} (p) (-im)\right)^{-1} \!\!S^{(0)} (p)\nonumber\\
& = \sum_{n=0}^{\infty} \left(S^{(0)}(p) (-im)\right)^{n} S^{(0)}(p)\nonumber\\
& = S^{(0)} (p) + S^{(0)} (-im) S^{(0)} (p) + \cdots,\label{freelsexpansion}
\end{align}
These are the analogs of \eqref{lseqn} and \eqref{lseqnexpansion} at zero temperature. Furthermore,  from the form of the massive propagator in \eqref{freepropagator} we note that
\begin{equation}
\frac{(i)^{n}}{n!} \frac{\partial^{n} S^{(m)} (p)}{\partial m^{n}} = \left(S^{(m)} (p)\right)^{n+1},\quad n\geq 1,\label{massidentity0}
\end{equation}
so that we can write
\begin{align}
S^{(m)} (p) & = \sum_{n=0}^{\infty} \frac{(m)^{n}}{n!} \frac{\partial^{n} S^{(m)}(p)}{\partial m^{n}}\Big|_{m=0}\nonumber\\
& = \sum_{n=0}^{\infty} \frac{(-im)^{n} (i)^{n}}{n!} \frac{\partial^{n} S^{(m)}(p)}{\partial m^{n}}\Big|_{m=0}\nonumber\\
&  = \sum_{n=0}^{\infty} (-im)^{n} \left(S^{(0)} (p)\right)^{n+1}\nonumber\\
& = \sum_{n=0}^{\infty}  \left(S^{(0)} (p) (-im)\right)^{n} S^{(0)}(p).
\end{align}
This shows that the expansion \eqref{freelsexpansion} is merely a Taylor expansion in the mass around $m=0$.  

At finite temperature, however, the propagators are $2\times 2$ matrices in the thermal space. For example, in the closed time path formalism the free fermion propagator (in momentum space in any dimension) has the form
\begin{equation}
S^{(m)}(p) = \begin{pmatrix}
S^{(m)}_{++} (p) & S^{(m)}_{+-}(p)\\
S^{(m)}_{-+}(p) & S^{(m)}_{--}(p)
\end{pmatrix},\label{closedtimepropagator}
\end{equation}
where the components are given by \cite{das} 
\begin{widetext}
\begin{align}
S^{(m)}_{++}(p) & = (p\!\!\!\slash +m)\left(\frac{i}{p^{2}-m^{2}+i\epsilon} - 2\pi n_{\rm F} (|p^{0}|) \delta (p^{2}-m^{2})\right),\quad
S^{(m)}_{+-}(p)  = 2\pi (p\!\!\!\slash +m)\left(\theta(-p^{0})-n_{\rm F}(|p^{0}|)\right)\delta(p^{2}-m^{2}),\nonumber\\
S^{(m)}_{-+}(p) & = 2\pi (p\!\!\!\slash +m)\left(\theta(p^{0})-n_{\rm F}(|p^{0}|)\right)\delta(p^{2}-m^{2}),\quad 
S^{(m)}_{--}(p)  = (p\!\!\!\slash +m)\left(-\frac{i}{p^{2}-m^{2}-i\epsilon} - 2\pi n_{\rm F} (|p^{0}|) \delta (p^{2}-m^{2})\right),\label{closedtimepropagator1}
\end{align}
\end{widetext}
where $n_{\rm F}(|p^{0}|)$ denotes the Fermi-Dirac distribution function. If we write the delta function in a regularized manner as
\begin{equation}
2\pi \delta (p^{2}-m^{2}) = \frac{i}{p^{2}-m^{2}+i\epsilon} - \frac{i}{p^{2}-m^{2}-i\epsilon},
\end{equation}
then it is straightforward to derive
\begin{align}
\lefteqn{\frac{\partial}{\partial m} (2\pi (p\!\!\!\slash +m) \delta (p^{2}-m^{2})) = 2\pi (p\!\!\!\slash +m)^{2}}\nonumber\\
&\quad\times\left(\frac{1}{p^{2}-m^{2}+i\epsilon} + \frac{1}{p^{2}-m^{2}-i\epsilon}\right)\delta (p^{2}-m^{2}).
\end{align}
Using this, it can now be shown that even for the $2\times 2$ matrix propagator \eqref{closedtimepropagator} a relation analogous to \eqref{massidentity0} holds in the form
\begin{equation}
\frac{(i)^{n}}{n!} \frac{\partial^{n} (S^{(m)}(p)\tau)}{\partial m^{n}} = (S^{(m)}(p)\tau)^{n+1},\label{massidentity}
\end{equation}
so that we can write
\begin{align}
S^{(m)}(p)\tau & = \sum_{n=0}^{\infty} \frac{(-im)^{n} (i)^{n}}{n!} \frac{\partial^{n} (S^{(m)}(p)\tau)}{\partial m^{n}}\Big|_{m=0}\nonumber\\
& = \sum_{n=0}^{\infty} (-im)^{n} (S^{(0)}(p)\tau)^{n+1},\nonumber\\
{\rm or,}\quad S^{(m)}(p) & = \sum_{n=0}^{\infty}  \left(S^{(0)}(p) (-im\tau)\right)^{n} S^{(0)}(p).\label{freematrixlsexpansion}
\end{align}
This is the analog of the relation \eqref{lseqnexpansion} for the free thermal propagator and is a consequence of the identity \eqref{massidentity} satisfied by the matrix propagator at finite temperature. Although we have demonstrated the identity \eqref{massidentity} only for closed time path, it holds for any finite temperature contour in the complex $t$-plane. (The relation \eqref{massidentity} has been derived for thermofield dynamics in \cite{umezawa} where it is called the mass-derivative formula.) The important thing to note from these discussions is that the mass expansion for the propagator works even at finite temperature provided one sums over the thermal indices of the internal mass insertion vertices. Furthermore, the massless propagator $\hat{S}^{(0)}$ in \eqref{lseqn} denotes the complete propagator including the background interactions to all orders. Since it has already been shown \cite{dasfrenkel, dasfrenkel1} that $\hat{S}^{(0)}$ coincides with the perturbative result order by order (namely, it satisfies the Lippmann-Schwinger equation to all orders), it follows now (using \eqref{freematrixlsexpansion}) that the propagator $\hat{S}^{(m)}$ also coincides with the perturbative expansion of the propagator order by order.

Finally, we note that the mass term which couples the left handed and the right handed modes has the effect of introducing diagonal elements into the propagator (in the Dirac space) which was off-diagonal to begin with (see \eqref{explicitmasslesspropagator}). This is easily seen from \eqref{lseqn} 
\begin{align}
\hat{S}^{(m)} & = (\mathbbm{1} + im \hat{S}^{(0)}\tau)^{-1} \hat{S}^{(0)}\nonumber\\
& = (\mathbbm{1} + m^{2} (\hat{S}^{(0)}\tau)^{2})^{-1} (\mathbbm{1} - im \hat{S}^{(0)}\tau) \hat{S}^{(0)}\nonumber\\
& =  (\mathbbm{1} + m^{2} (\hat{S}^{(0)}\tau)^{2})^{-1} (\hat{S}^{(0)} - im \hat{S}^{(0)}\tau\hat{S}^{(0)})\nonumber\\
& = \hat{S}_{\rm D} + \hat{S}_{\rm O},\label{diagonal-offdiagonal}
\end{align}
where we have used the fact that an even number of $\hat{S}^{(0)}$ leads to a diagonal structure while an odd number of the massless propagator is off-diagonal (in the Dirac space). It also follows from \eqref{diagonal-offdiagonal} that the diagonal elements would arise from an odd number of mass insertions while the off-diagonal elements would involve an even number of mass insertions. When $m=0$, the diagonal elements vanish while the off-diagonal elements reduce to $\hat{S}^{(0)}$.

\subsection{Effective action}

With the determination of the complete propagator for the massive fermion, we are now in a position to determine the effective action. Even though \eqref{propagator0} gives the correct relation between the effective action and the propagator at zero temperature, since in the matrix formalism of closed time path at finite temperature, the number of fields is doubled and the mass term on the $C_{-}$ branch comes with an opposite sign, the correct relation in this case is given by
\begin{equation}
\frac{\partial \Gamma_{\rm eff}^{(m)}}{\partial m} = {\rm Tr}\ \tau \hat{S}^{(m)} = {\rm Tr}\ \hat{S}^{(m)} \tau,\label{propagatorfinal}
\end{equation}
where $\tau$ is the thermal matrix defined in \eqref{thermalmatrices} and ``Tr" here denotes a trace over the Dirac indices as well as over the thermal indices and also a trace over a complete basis. Noting the form of the complete massive propagator in \eqref{lseqn}, we see that \eqref{propagatorfinal} can be integrated to give
\begin{equation}
\Gamma_{\rm eff}^{(m)} = -i\, {\rm Tr}\ \ln \left(\mathbbm{1} + im \hat{S}^{(0)} \tau\right),\label{effectivemaction}
\end{equation}
up to mass independent as well as normalization terms. 

It is clear that when $m=0$, the effective action in \eqref{effectivemaction} formally vanishes. This determines the mass independent term to coincide with the massless effective action $\Gamma_{\rm eff}^{(0)}$ determined in \cite{dasfrenkel, dasfrenkel1,adilson}. The massless effective action is already determined with the normalization $\Gamma_{\rm eff}^{(0)} (e=0) = 0$. Thus, we can determine the complete normalized effective action for massive QED to be
\begin{equation}
\Gamma_{\rm eff} = \Gamma_{\rm eff}^{(0)} + \Gamma_{\rm eff}^{(m)} - \Gamma_{\rm eff}^{(m)} (e=0),\label{completeeffectiveaction}
\end{equation} 
such that
\begin{equation}
\Gamma_{\rm eff} (e=0) = 0,\quad \Gamma_{\rm eff} (m=0) = \Gamma_{\rm eff}^{(0)}.\label{normalization}
\end{equation}
The form of the complete effective action \eqref{completeeffectiveaction} is consistent with the expectation from the definition of the effective action. Namely, the definition of the effective action \eqref{effaction} in the case of thermal doublet fields (see, for example,  \eqref{matrixpropagator} and \eqref{masslessmatrixpropagator}) leads to (the effective action is not normalized)
\begin{align}
\Gamma_{\rm eff} & = -i {\rm Tr}\ \ln \left(\gamma^{\mu} (i\partial_{\mu} - e \hat{A}_{\mu}) - m\tau\right)\nonumber\\
& = - i {\rm Tr}\ \ln \left(i (\hat{S}^{(0)})^{-1} (\mathbbm{1} + im \hat{S}^{(0)}\tau)\right)\nonumber\\
& = - i\left[{\rm Tr}\ \ln i(\hat{S}^{(0)})^{-1} + {\rm Tr}\ \ln (\mathbbm{1} + im \hat{S}^{(0)}\tau)\right]\nonumber\\
& = \Gamma_{\rm eff}^{(0)} + \Gamma_{\rm eff}^{(m)}.\label{effectivemaction1}
\end{align}
We note here that this result (\eqref{effectivemaction} or \eqref{effectivemaction1}) holds in any dimension. However, in $1+1$ dimensions we have the advantage that we have already determined $\hat{S}^{(0)}$ \cite{dasfrenkel, dasfrenkel1}.

Let us next analyze the structure of $\Gamma_{\rm eff}^{(m)}$ in some detail. We recall that since $\hat{S}^{(0)}$ is off-diagonal in the Dirac space (see \eqref{explicitmasslesspropagator}) and since the definition of $\Gamma_{\rm eff}^{(m)}$ in \eqref{effectivemaction} involves a Dirac trace, the odd powers of $\hat{S}^{(0)}$ in the expansion of the logarithm vanish and we can effectively write
\begin{align}
\Gamma_{\rm eff}^{(m)} & = -i \sum_{n=1}^{\infty} {\rm Tr}\ (-1)^{n+1}\, \frac{(im\hat{S}^{(0)} \tau)^{n}}{n}\nonumber\\
& =  -i \sum_{n=1}^{\infty} {\rm Tr}\ (-1)^{2n+1}\, \frac{(im\hat{S}^{(0)} \tau)^{2n}}{2n}\nonumber\\
& = - \frac{i}{2} \sum_{n=1}^{\infty} {\rm Tr}\ (-1)^{n+1}\, \frac{(m^{2}(\hat{S}^{(0)} \tau)^{2})^{n}}{n}\nonumber\\
& = - \frac{i}{2}\, {\rm Tr}\ \ln \left(\mathbbm{1} + m^{2} (\hat{S}^{(0)}\tau)^{2}\right)\nonumber\\
& = - \frac{i}{2}\, {\rm Tr}\ \ln \left(\mathbbm{1} -  (\hat{S}^{(0)} (-im\tau))^{2}\right).\label{simplified}
\end{align}
This result that the effective action depends effectively on $m^{2}$ can be understood from the fact that the generating functional for the Dirac theory does not change under a field redefinition
\begin{equation}
\psi \rightarrow \gamma_{5} \psi,\quad \overline{\psi} \rightarrow - \overline{\psi} \gamma_{5},
\end{equation}
which translates into the invariance of the effective action under
\begin{equation}
m \rightarrow - m.
\end{equation}
This can also be seen directly from the definition \eqref{effectivemaction}
\begin{align}
\Gamma_{\rm eff}^{(m)} & = - i\, {\rm Tr}\ \ln \left(\mathbbm{1} + im \hat{S}^{(0)} \tau\right)\nonumber\\
& = - i\, {\rm Tr}\ \ln \gamma_{5}^{2}\left(\mathbbm{1} + im \hat{S}^{(0)} \tau\right)\nonumber\\
& = - i\, {\rm Tr}\ \ln \gamma_{5}\left(\mathbbm{1} - im \hat{S}^{(0)} \tau\right)\gamma_{5}\nonumber\\
& = - i\, {\rm Tr}\ \ln \left(\mathbbm{1} - im \hat{S}^{(0)} \tau\right) = \Gamma_{\rm eff}^{(-m)},
\end{align}
where we have used the cyclicity of trace (under an expansion of the logarithm) as well as the fact that $\gamma_{5}$ anti-commutes with $\gamma^{\mu}$. As a result, we have
\begin{align}
\Gamma_{\rm eff}^{(m)} & = - \frac{i}{2}\, {\rm Tr}\left[\ln \left(\mathbbm{1} + im \hat{S}^{(0)} \tau\right) + \ln \left(\mathbbm{1} - im \hat{S}^{(0)} \tau\right)\right]\nonumber\\
& = - \frac{i}{2}\, {\rm Tr}\ \ln \left(\mathbbm{1} + m^{2}  (\hat{S}^{(0)} \tau)^{2}\right),
\end{align} 
which coincides with \eqref{simplified}.

\begin{figure}[ht!]
\begin{center}
\includegraphics[scale=.8]{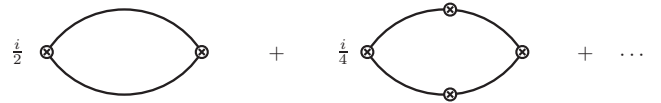}
\end{center}
\caption{Diagrammatic expansion of $\Gamma_{\rm eff}^{(m)}$. The solid lines represent $\hat{S}^{(0)}$ while a vertex $\otimes$ denotes a mass insertion $(-im\tau)$ and the ``Tr" operation is understood.}
\label{3}
\end{figure}

Following Schwinger, we can give an integral (``proper time") representation for $\Gamma_{\rm eff}^{(m)}$ in \eqref{simplified} as (see \eqref{effaction1})
\begin{equation}
\Gamma_{\rm eff}^{(m)} =  \lim_{\nu\rightarrow 0}\, \frac{i}{2} \int\limits_{0}^{\infty} \frac{ds}{s^{1-\nu}}\, {\rm Tr}\,e^{- s}\, e^{-s m^{2} (\hat{S}^{(0)} \tau)^{2}},
\end{equation}
where we have denoted the ``proper time" parameter by $s$ to avoid confusion with the thermal matrix $\tau$. From a calculational point, however, the effective action \eqref{simplified} is best viewed in terms of an expansion of the logarithm (we will discuss the regime of validity of such an expansion shortly)
\begin{equation}
\Gamma_{\rm eff}^{(m)}  = \frac{i}{2} {\rm Tr} \left(\hat{S}^{(0)} (-im\tau)\right)^{2} + \frac{i}{4} {\rm Tr}\left(\hat{S}^{(0)} (-im\tau)\right)^{4} + \cdots ,\label{massexpansion}
\end{equation}
which can be given a diagrammatic representation as shown in Fig. \ref{3}.
Such an expansion in powers of mass insertions can be thought of as a mass perturbation of the effective action where $\hat{S}^{(0)}$ represents the complete massless propagator including background interactions to all orders. As we have shown in \cite{dasfrenkel, dasfrenkel1} this propagator coincides with the perturbation expansion in a massless theory order by order. In a conventional perturbation expansion, on the other hand, one starts with the free massive fermion propagator and perturbs in the background interactions. Using \eqref{freematrixlsexpansion} it can be shown that the contributions from the effective action $\Gamma_{\rm eff}^{(m)}$ in \eqref{simplified} coincide with the mass corrections following from the conventional perturbation in the massive theory order by order in $m$ and $e$. 

Given the complete form of the effective action \eqref{completeeffectiveaction}, we can now discuss various of its features. We already know \cite{dasfrenkel, dasfrenkel1, adilson} that the temperature dependent effective action for the massless theory $\Gamma_{\rm eff}^{(0)}$ is purely imaginary,
\begin{equation}
\left(\Gamma_{\rm eff}^{(0)}\right)^{*} = - \Gamma_{\rm eff}^{(0)}.\label{masslessimaginary}
\end{equation}
This is easily seen by noting that the effective action for the Schwinger model consists of only even number of photon fields (by charge conjugation invariance). From \eqref{raction} we note that since such terms coming from $\hat{\cal O}_{+}$ are all real, it is the overall factor of ``$i$" that makes the effective action purely imaginary. However, the mass corrections coming from $\Gamma_{\rm eff}^{(m)}$ are in general complex. This follows from the fact that the (matrix) thermal propagator $\hat{S}^{(0)}$ is, in general, complex (see, for example, \eqref{closedtimepropagator1}). Therefore, we can write
\begin{align}
{\rm Im}\, \Gamma_{\rm eff}^{(m)} & = - \frac{i}{2}\left(\Gamma_{\rm eff}^{(m)} - (\Gamma_{\rm eff}^{(m)})^{*}\right)\nonumber\\
 & = - \frac{1}{4}\, {\rm Tr}\left[\ln \left(\mathbbm{1} + m^{2}  (\hat{S}^{(0)} \tau)^{2}\right)\right.\nonumber\\  & \qquad\qquad\quad \left.+ \ln \left(\mathbbm{1} + m^{2}  ((\hat{S}^{(0)})^{*} \tau)^{2}\right)\right].
 \end{align}

\begin{figure}[ht!]
\begin{center}
\includegraphics[scale=.8]{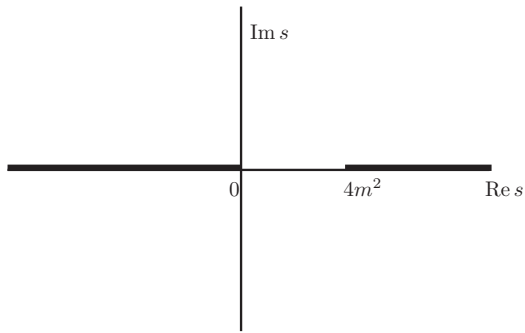}
\end{center}
\caption{The two branch cuts in the two point function in massive QED at finite temperature. Here $s\equiv p^{2}$ where $p^{\mu}$ denotes the external momentum. The branch cut for $s\geq 4m^{2}$ corresponds to pair production while the other for $s\leq 0$ represents the effects due to thermal scattering.}
\label{4}
\end{figure}

At zero temperature the imaginary part of the effective action is related to the probability of vacuum decay through pair production. At finite temperature, on the other hand, there are other channels of reaction possible. For example, a particle travelling through a thermal medium can absorb a physical (on-shell) particle from the medium and emit an on-shell particle into the medium. Such processes, commonly known as thermal scattering, also lead to a destabilization of the vacuum (this would become clear in the explicit calculation presented in the next section) and the imaginary part of the effective action at finite temperature should be thought of as the probability of vacuum decay through all such processes. In fact, the new thermal processes introduce new branch cuts into amplitudes at finite temperature. For example, in massive QED there are two branch cuts in the two point function at finite temperature as shown in Fig. \ref{4}.
Therefore, the imaginary part of the two point function in massive QED at finite temperature can be written, in general, as ($s\equiv p^{2}$)
\begin{equation}
{\rm Im}\, \Pi (p,m) = A\, \theta(s-4m^{2}) + B\, \theta (-s),\label{branchcut}
\end{equation}
where $A,B$ depend on external momentum as well as mass. Equation \eqref{branchcut} brings out an interesting aspect of thermal amplitudes, namely, the two point amplitude is non-analytic at $s=0, m=0$. Explicitly, we see that 
\begin{align}
& \lim_{s\rightarrow 0} \lim_{m\rightarrow 0}\, {\rm Im}\,\Pi (p,m) = \frac{1}{2}\left(A + B\right)_{m=0=s},\nonumber\\
& \lim_{m\rightarrow 0} \lim_{s\rightarrow 0}\, {\rm Im}\,\Pi (p,m) = \frac{1}{2}\,B\big|_{m=0=s},\label{discontinuity}
\end{align} 
where we have used $\theta (0)=\frac{1}{2}$. We note that for a massless theory ($m=0$), the two branch cuts in Fig. \ref{4} merge into a single branch cut given by $-\infty< s<\infty$ and, in this case, the pair production and thermal scattering are described respectively by the regimes $s\geq 0$ and $s\leq 0$.

\begin{figure}[ht!]
\begin{center}
\includegraphics[scale=.7]{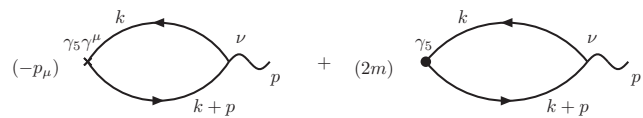}
\end{center}
\caption{The two diagrams corresponding to \eqref{chiralWI} at the two point level.}
\label{5}
\end{figure}

As in the Schwinger model, there is no finite temperature corrections to the axial anomaly in the massive QED. This is most easily seen by noting that the classical (tree level) Ward identity associated with an  infinitesimal chiral transformation in the massive theory is given by
\begin{equation}
\partial_{\mu} J_{5}^{\mu} (x) + 2im J_{5} (x) = 0,\label{chiralWI}
\end{equation}
where
\begin{equation}
J_{5}^{\mu} = \overline{\psi} \gamma_{5}\gamma^{\mu}\psi,\quad J_{5} = \overline{\psi}\gamma_{5}\psi.
\end{equation}
At one loop level, the left hand side of Ward identity \eqref{chiralWI} would lead to the two point functions described by the Feynman diagrams shown in Fig. \ref{5}.
The finite temperature contribution for the sum of these two graphs is proportional to (say for the thermal $++$ amplitudes, see also \eqref{closedtimepropagator1})
\begin{widetext}
\begin{align}
&  \int d^{2}k\left[(k^{2}-m^{2}) \epsilon^{\mu\nu} (k+p)_{\mu} - ((k+p)^{2}-m^{2})\epsilon^{\mu\nu} k_{\mu}\right]\nonumber\\
&\qquad\times n_{\rm F} (|k^{0}|)\left(\frac{i}{(k+p)^{2}-m^{2}+i\epsilon} - \pi  n_{\rm F} (|k^{0}+p^{0}|) \delta ((k+p)^{2}-m^{2})\right)  \delta(k^{2}-m^{2}) =  0,\label{anomaly}
\end{align}
\end{widetext}
where $\epsilon^{\mu\nu}$ denotes the two dimensional Levi-Civita tensor and we use the convention $\epsilon^{01}=1$. The terms quadratic in the distribution functions in \eqref{anomaly} vanish because of the delta function constraints while those linear in the distribution function vanish because of the delta function constraint as well as by anti-symmetry. This shows that the classical (tree level) chiral Ward identity \eqref{chiralWI} holds at finite temperature as well and, consequently, there is no finite temperature contribution to the anomaly in the massive QED as in the Schwinger model.

In the case of the Schwinger model, it has been shown \cite{dasfrenkel, dasfrenkel1, adilson} that the retarded/advanced amplitudes vanish at finite temperature. In contrast, in massive QED, the retarded/advanced amplitudes do not vanish at finite temperature and have the correct behavior expected in  thermal field theory. This will be discussed in some detail in the next section.

As we have noted earlier, from a calculational point of view the effective action $\Gamma_{\rm eff}^{(m)}$ in \eqref{simplified} can be best understood as an expansion \eqref{massexpansion}. This would correspond to a mass expansion (perturbation) around a massless theory. At zero temperature such an expansion in a $1+1$ dimensional theory is mildly infrared divergent and there are well known methods \cite{coleman} for handling this problem so that it is not really a difficulty. The second feature that arises at zero temperature in one loop amplitudes is that the amplitudes develop a dependence on $m^{2} \ln m^{2}$ (say at order $m^{2}$ of the expansion) so that the expansion ceases to be a true expansion in powers of mass $m$.  At finite temperature, on the other hand, the second problem disappears in the sense that the temperature dependent amplitudes also develop a logarithmic mass dependence which exactly cancels the zero temperature term. Therefore, the expansion \eqref{massexpansion} is truly a mass expansion at finite temperature. However, the problem of infrared divergence is much more severe at finite temperature. We recall that there are two sources of infrared divergence at finite temperature. First, the distribution function itself can have a divergent infrared behavior in a bosonic theory, but the fermion distribution function which we are dealing with is free from this problem. However, there is a second source of infrared divergence that is common to all theories which arises from the fact that the thermal propagators (see \eqref{closedtimepropagator1}) have  $\delta$-function terms which describe on-shell particles in the thermal medium. In a finite temperature diagram describing a thermal amplitude, therefore, there will be one or more on-shell particles in the loop. Such diagrams can become infrared divergent when the external (massless) particle is also on-shell. This is a physical divergence which cannot be taken care of by the zero temperature techniques. As a result, the mass expansion \eqref{massexpansion} becomes meaningful only if (say, for the two point amplitude) $|p^{2}| \gg m^{2}$, where $p^{\mu}$ denotes the momentum of the external particle. We will discuss this in more detail in the next section with an explicit calculation.

\section{Explicit calculation}

\begin{figure}[ht!]
\begin{center}
\includegraphics[scale=.8]{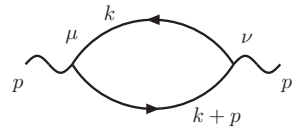}
\end{center}
\caption{The two point amplitude for the photon on the $C_{+}$ thermal branch. The solid internal lines denote $S_{++}^{(m)}$ given in \eqref{closedtimepropagator1}.}
\label{6}
\end{figure}

In this section, we will calculate explicitly the thermal two point function on the thermal $C_{+}$ branch to support various observations made in the last section. We have already argued that the expansion following from \eqref{completeeffectiveaction} coincides exactly with that we have in perturbation theory. The two expansions are really just the opposite of each other in the sense that in the effective action \eqref{completeeffectiveaction}, $\hat{S}^{(0)}$ contains the gauge interactions to all orders and $\Gamma_{\rm eff}^{(m)}$ leads to an expansion (perturbation) in powers of $m$. In usual perturbation theory (for amplitudes on the $C_{+}$ branch), on the other hand, $\hat{S}_{++}^{(m)}$ (see \eqref{closedtimepropagator1}) contains the complete dependence on $m$ (at the tree level) while the perturbation is in powers of the coupling constant $e$. The two coincide order by order in $(m,e)$ and,  therefore, we look at the two point amplitude in perturbation theory (on the $C_{+}$ thermal branch) which is described by the Feynman diagram shown in Fig. \ref{6}.
The temperature dependent part of the amplitude in Fig. \ref{6} is given by
\begin{widetext}
\begin{align}
- i \Pi^{\mu\nu} (p) & = -\frac{2ie^{2}}{\pi}\int d^{2}k\left(k^{\mu} (k+p)^{\nu} + k^{\nu}(k+p)^{\mu} - \eta^{\mu\nu} (k\cdot (k+p) - m^{2})\right)\nonumber\\
& \qquad\times \left(\frac{1}{(k+p)^{2} - m^{2}+i\epsilon} + i\pi n_{\rm F} (|k^{0}+p^{0}|) \delta ((k+p)^{2}-m^{2})\right) n_{\rm F} (|k^{0}|) \delta (k^{2}-m^{2}).\label{polarization}
\end{align}
\end{widetext}

The two point amplitude for the photon is transverse to the external momentum and, in general, there are two independent transverse structures at finite temperature \cite{das}. However, in $1+1$ dimensions a simplification occurs and one of the transverse structures, which is transverse to both the momentum as well as the velocity of the heat bath, vanishes. The reason for this simplification lies in the fact that in $1+1$ dimension the transverse direction to a given vector $A^{\mu}$ is uniquely given by (up to a scaling) $\epsilon^{\mu\nu} A_{\nu}$ (independent of any other vector) where $\epsilon^{\mu\nu}$ denotes the Levi-Civita tensor  \cite{hott}. (An alternative way of understanding why the second transverse structure vanishes is that since a vector in $1+1$ dimensions has only two components, it cannot be simultaneously orthogonal to two vectors unless the two are collinear.)  For example, we note that if $u^{\mu}$ denotes the velocity of the heat bath, we can define two Lorentz invariant quantities from the momentum $p^{\mu}$ as 
\begin{equation}
\omega = (p\cdot u),\quad p' = \epsilon^{\mu\nu} p_{\mu}u_{\nu},\label{lorentzinvariants}
\end{equation}
so that we can write
\begin{equation}
p^{\mu} = \omega u^{\mu} - p' \epsilon^{\mu\nu} u_{\nu}.\label{pmu}
\end{equation}
We note that in the rest frame of the heat bath ($u^{\mu}=(1,0)$), we have $\omega = p_{0}, p' = -p_{1}$. With these, we can define the component of the velocity $u^{\mu}$ orthogonal to $p^{\mu}$ as
\begin{equation}
\bar{u}^{\mu} = u^{\mu} - \frac{\omega}{p'}\,\epsilon^{\mu\nu} u_{\nu} = - \frac{1}{p'}\, \epsilon^{\mu\nu} p_{\nu},\label{ubar}
\end{equation}
where we have used \eqref{pmu} in the last step. It follows now that
\begin{align}
\bar{u}^{\mu} \bar{u}^{\nu} & = \left(-\frac{1}{p'}\,\epsilon^{\mu\lambda} p_{\lambda}\right)\left(-\frac{1}{p'}\,\epsilon^{\nu\rho} p_{\rho}\right)\nonumber\\
 & = - \frac{p^{2}}{(p')^{2}}\left(\eta^{\mu\nu} - \frac{p^{\mu}p^{\nu}}{p^{2}}\right),
\end{align}
which corresponds to the conventional transverse structure up to a scaling. The vector $\bar{u}^{\mu}$ can also be expressed in terms of the light-cone components of the velocity as
\begin{equation}
\bar{u}^{\mu} = -\frac{1}{2p'}\left((\omega - p') u^{\mu}_{-} - (\omega+p') u^{\mu}_{+}\right),\label{light-cone}
\end{equation}
where the light-cone components of the velocity vector are defined as
\begin{equation}
u^{\mu}_{\pm} = \left(\eta^{\mu\nu} \mp \epsilon^{\mu\nu}\right)u_{\nu},
\end{equation}
and in the rest frame of the heat bath we have
\begin{equation}
\omega \pm p' = p_{0} \mp p_{1} = p_{\mp} = p\cdot u_{\mp}.\label{light-conep}
\end{equation}

The other interesting consequence of $1+1$ dimensions is that the transverse structure of the polarization tensor can be factored out of the integral in \eqref{polarization} \cite{hott}. The remaining integral can be written in several equivalent ways each differing from the other by a quantity that vanishes upon integration. Each of these may be of interest and use in a particular study. For example, for our purposes we can write the polarization tensor in \eqref{polarization} as
\begin{widetext}
\begin{equation}
\Pi^{\mu\nu} (p) = \frac{4e^{2} m^{2}}{\pi} \frac{\bar{u}^{\mu}\bar{u}^{\nu}}{(\bar{u}\cdot \bar{u})} \int d^{2}k \left(\frac{1}{(k+p)^{2} - m^{2}+i\epsilon} + i\pi n_{\rm F} (|k^{0}+p^{0}|) \delta ((k+p)^{2}-m^{2})\right) n_{\rm F} (|k^{0}|) \delta (k^{2}-m^{2}),\label{polarization1}
\end{equation}
\end{widetext}
which is useful for working with the massive theory. However, this structure is not convenient for taking the massless limit for which an alternative expression is more useful (such a structure naturally arises in the Schwinger model \cite{adilson}). We note from \eqref{polarization1} that we can write 
\begin{equation}
\Pi^{\mu\nu} (p,m) = \frac{\bar{u}^{\mu}\bar{u}^{\nu}}{\bar{u}\cdot \bar{u}}\, \Pi (p,m),\label{Pi}
\end{equation}
where $\frac{\bar{u}^{\mu}\bar{u}^{\nu}}{\bar{u}\cdot \bar{u}}$ is the conventional transverse projection operator \cite{das,hott} which has factored out of the integral. 

Let us study the two point amplitude separately in the limits $m\rightarrow 0$ and $p^{2}\rightarrow 0$. Since the projection operator $\frac{\bar{u}^{\mu}\bar{u}^{\nu}}{\bar{u}\cdot \bar{u}}$ is not well defined in the limit $p^{2}\rightarrow 0$, we look at the amplitude as a whole for this purpose. For example, it follows from \eqref{polarization1} that in the massless limit 
\begin{equation}
\Pi^{\mu\nu} (p^{2}\neq 0, m\rightarrow 0) \sim \frac{m^{2}}{(p^{2})^{2}} \bar{u}^{\mu} \bar{u}^{\nu}  \rightarrow  0,\label{massless}
\end{equation}
so that the result coincides with that of the massless theory \cite{dasfrenkel,dasfrenkel1,adilson}. On the other hand, in the limit $p^{2}=0$, the integral in \eqref{polarization1} leads to
\begin{equation}
{\rm Im}\, \Pi^{\mu\nu} (p^{2}=0, m) = 0.
\end{equation}
This can be physically understood by noting from \eqref{polarization1} that when $p^{2}=0$, the imaginary part of $\Pi^{\mu\nu}$ arises when the external as well as both the internal lines in Fig. \ref{6} are on-shell. However, since the internal lines correspond to massive fermions, this is not kinematically allowed. For $T\gg m$, the real part of the amplitude \eqref{polarization1} can be evaluated in the rest frame of the heat bath to give
\begin{equation}
{\rm Re}\, \Pi^{\mu\nu} (p^{2}=0,m) = \bar{u}^{\mu}\bar{u}^{\nu}\!\left(\frac{\pi e^{2}}{3}\, \frac{T^{2}}{m^{2}} + O \left(\ln (T^{2}/m^{2})\right)\!\right)\!.\label{infrared}
\end{equation}
There are several things to note from these results. For example, \eqref{massless} and \eqref{infrared} show that  ${\rm Re}\, \Pi^{\mu\nu}(p,m)$ is non-analytic at $m=0, s=p^{2}=0$. This non-analyticity has its origin in the branch cuts that we have alluded to in \eqref{branchcut} and \eqref{discontinuity}. The other interesting point to note from \eqref{infrared} is that the high temperature behavior of the two point amplitude is superleading which is unexpected from a naive dimensional analysis. This can, in fact, be traced to the strong infrared divergence (in the limit $m\rightarrow 0$) at finite temperature that we have pointed out earlier.  

It is clear from the explicit results in \eqref{massless} and \eqref{infrared} that the real part of $\Pi^{\mu\nu}$ can be modeled to have the form
\begin{equation}
{\rm Re}\,\Pi^{\mu\nu} (p,m) \sim \frac{m^{2}}{a (m^{2})^{2} + b (p^{2})^{2}}\ \bar{u}^{\mu}\bar{u}^{\nu},\label{model}
\end{equation}
where $a,b$ are temperature dependent constant parameters. This would be consistent with the behaviors \eqref{massless} and \eqref{infrared} in the appropriate limits and the denominator in \eqref{model} shows the expected non-analyticity at $m = 0 = p^{2}$. It also makes clear that an expansion of \eqref{model} in powers of mass would be meaningful only if $|p^{2}| \gg m^{2}$. This reflects in the fact that a mass expansion of $\Gamma_{\rm eff}^{(m)}$ is meaningful only in this regime as we have pointed out earlier. In the regime $|p^{2}| \gg m^{2}$, the two point function \eqref{polarization} can be determined exactly. The  real and the imaginary parts of $\Pi$ in this region are given by (the projection operator is well behaved in this regime) 
\begin{widetext}
\begin{align}
{\rm Re}\, \Pi (p,m) & = \frac{2e^{2}m^{2}}{\pi p^{2}}\left(\ln |\frac{p_{+}}{2m}| +  {\rm Re}\left[\zeta (1, \frac{1}{2} + \frac{ip_{+}}{4\pi T}) - \zeta(1,\frac{1}{2})\right] + (p_{+}\rightarrow -p_{-})\right),\label{zeta}\\
{\rm Im}\, \Pi (p,m) & = - \frac{2e^{2} m^{2}\,{\rm sgn} (p^{2})}{p^{2}}\left(n_{\rm F} (|\frac{p_{+}}{2}|) \big(1 - n_{\rm F} (|\frac{p_{-}}{2}|)\big) + p_{+}\rightarrow - p_{-}\right).\label{exact}
\end{align}
\end{widetext}
where $\zeta (1, q)$ in \eqref{zeta} represents the Riemann zeta function \cite{GR}. We note here that individually, each of the zeta functions $\zeta (1,q)$ in \eqref{zeta} diverges logarithmically, but their difference is well behaved so that the expression as a whole is well defined. (It is also worth pointing out here that since we are assuming $|p^{2}|\gg m^{2}$, the exact calculation leading to \eqref{zeta} and \eqref{exact} was carried out by neglecting the mass $m$ in the integrand \eqref{polarization1} where ever possible.) We also note that ${\rm Re}\, \Pi$ in \eqref{zeta} supports the behavior in \eqref{model} for $|p^{2}| \gg m^{2}$. Furthermore, we can compare the imaginary part of $\Pi$ in \eqref{exact} with \eqref{branchcut} and see that it has contributions from both $p^{2} > 0$ and $p^{2} < 0$. The contributions from both these regimes would contribute to a decay of the vacuum as we have emphasized earlier. From \eqref{zeta} we note that since the real part of the amplitude is non-vanishing and the real part of the Feynman amplitude is related to that of the retarded amplitude as \cite{das}
\begin{equation}
{\rm Re}\, \Pi = {\rm Re}\, \Pi_{\rm ret},
\end{equation}
the retarded amplitudes do not vanish in the massive theory unlike in the Schwinger model (however, they do vanish in the $m\rightarrow 0$ limit \cite{retarded}). We have explicitly verified this as well as the expected dispersion relations \cite{das} at finite temperature.

\section{Summary}

In this paper we have extended our earlier proposal \cite{dasfrenkel, dasfrenkel1} for evaluating the thermal effective actions to the case of $1+1$ dimensional massive QED. We have determined the exact fermion propagator of the theory which then leads to the thermal effective action. This effective action systematically generates all the one loop Feynman amplitudes in thermal perturbation theory. We have discussed various features of the effective action including its imaginary part. The general observations regarding the effective action have been further strengthened through an explicit calculation of the quadratic effective action at finite temperature.

\bigskip

\noindent{\bf Acknowledgments}
\medskip

This work was supported in part  by US DOE Grant number DE-FG 02-91ER40685,  by CNPq and 
FAPESP (Brazil).

\end{document}